\documentstyle[12pt,epsf,twoside,fleqn,espcrc1]{article}

\newcommand{\be}{\begin{eqnarray}}
\newcommand{\ee}{\end{eqnarray}}

\newcommand{\AmS}{{\protect\the\textfont2
  A\kern-.1667em\lower.5ex\hbox{M}\kern-.125emS}}

\title{Shadowing Effects on Particle and
Transverse Energy Production\thanks{This work was supported
in part by the Director, Office of Energy Research, Division of Nuclear Physics
of the Office of High Energy and Nuclear Physics of the U. S.
Department of Energy under Contract No. DE-AC03-76SF0098.}}

\author{V. Emel'yanov$^{\rm \tiny a}$, A. Khodinov\address{Moscow State 
Engineering Physics Institute (Technical
University), Kashirskoe Ave. 31, Moscow, 115409, Russia}, S. R. 
Klein\address{Lawrence Berkeley National 
Laboratory, Berkeley, CA 94720, USA} and 
R. Vogt$^{\rm
\tiny b,}$\address{Physics Department, University of 
California, Davis, CA 95616, USA}}

\begin{document}

\maketitle

\begin{abstract}
The effect of shadowing on the early state of 
ultrarelativistic heavy ion collisions and
transverse energy production is discussed.  Results
are presented for RHIC Au+Au collisions at $\sqrt{s_{NN}} = 200$ GeV and LHC
Pb+Pb collisions at $\sqrt{s_{NN}} = 5.5$ TeV.
\end{abstract}\\

The proton and neutron
structure functions are modified in the nuclear environment \cite{Arn}, 
referred
to here as shadowing.  In most shadowing models, such as gluon recombination,
the structure function modifications should be correlated
with the local nuclear density.  
The E745 experiment studied the spatial distribution of structure functions
with $\nu N$ interactions in emulsion \cite{E745}.  
They found evidence of a spatial dependence
but could not determine the form.
The spatial dependence of shadowing is reflected in particle
production as a function of impact parameter, $b$, which may be
inferred from the total transverse energy, $E_T$, produced in a heavy
ion collision \cite{us}. 

At RHIC and LHC perturbative QCD processes are expected to be an important
component of the total particle production.
In particular, at early times, $\tau_i \sim
1/p_T\leq 1/p_0\sim 0.1$ fm for $p_0 \sim 2$ GeV, semihard production
of low $p_T$ minijets sets the stage for further evolution \cite{EG}.

The average transverse energy of minijet production
is proportional to the initial energy density and
is the hard scattering contribution to the average total transverse energy.
It is calculated within a specific detector acceptance $\epsilon(y)$.
At leading order, the 
average transverse energy for a given minijet flavor, $f$, as
a function of impact parameter is \begin{eqnarray}
\overline E_T^f(b) = \frac{d\sigma^f\langle E_T^f \rangle}{d^2b} \!\!\! & = & 
\!\!\! K_{\rm jet} \int d^2r dp_T^2 dy_2 dy dz dz'
\sum_{{ij=}\atop{\langle kl \rangle}}
x_1F_i^A(x_1,p_T^2,\vec{r},z) x_2 F_j^B(x_2,p_T^2,\vec{b} - \vec{r},z')
\nonumber \\
&  & \mbox{} \times \frac{ \epsilon(y) p_T}{1 + \delta_{kl}}
\left[\delta_{fk} {d\hat\sigma\over d\hat t}^{ij\rightarrow kl}(\hat t, \hat u)
+ \delta_{fl} {d\hat\sigma\over d\hat t}^{ij\rightarrow kl}(\hat u, \hat t)
\right] \, \, . \label{etmom} \end{eqnarray}
Identical particles in
the final state are accounted for by the factor $1/(1 + \delta_{kl})$.  
The ratio of the NLO to LO jet cross sections, $K_{\rm jet}$,
indicates the size of
the NLO corrections. A conservative $K_{\rm jet}=1$ \cite{EKinit} 
gives a lower limit on minijet
production.  The cutoff $p_0$ is the lowest $p_T$ scale at
which perturbative QCD is valid. 

The nuclear parton densities, $F_i^A$, are assumed be the product of
the nucleon density in the nucleus $\rho_A(\vec{r}, z)$, 
the nucleon parton density
$f_i^N(x,Q^2)$, and a shadowing function $S^i(A,x,Q^2,\vec{r},z)$
where $A$ is the atomic mass number, $x$ is the parton momentum fraction,
$Q^2$ is the interaction scale, and $\vec{r}$ and $z$ are the transverse and
longitudinal location of the parton so that
$F_i^A(x,Q^2,\vec{r},z) = \rho_A(\vec{r}, z) S^i(A,x,Q^2,\vec{r},z)
f_i^N(x,Q^2).$ 
In the absence of nuclear modifications,
$S^i(A,x,Q^2,\vec{r},z)\equiv1$. If the nuclear structure functions are
homogeneous, then the spatial effects factorize and $\overline E_T^f(b,
p_0)$ is proportional to the number of nucleon-nucleon collisions, $T_{AB}(b)$.
We have calculated our results with GRV
94 LO \cite{grv94} and MRST LO \cite{mrsg} parton distributions.  The GRV low
$x$ gluon density is larger than the MRST and produces strikingly
different results at LHC while the RHIC results are rather similar.
We use
three different parameterizations of shadowing, all based on nuclear deep
inelastic scattering.  The first
parameterization, $S_1(A,x)$, treats quarks, antiquarks, and gluons
identically without $Q^2$ evolution \cite{EQC}. The others evolve with
$Q^2$ and conserve baryon number and total momentum.  The second,
$S_2^i(A,x,Q^2)$, modifies the valence quarks, sea quarks and gluons
separately and includes $Q^2$ evolution for $Q_0=2 < Q < 10$ GeV
\cite{KJE}.  The third, $S_3^i(A,x,Q^2)$, evolves each parton type separately
above $Q_0 = 1.5$ GeV \cite{EKRS3}.  
We have studied the spatial dependence of shadowing using two
parameterizations. 
If shadowing is proportional to the local
density \cite{us,firstprl},
\begin{eqnarray}
S^i_{\rm WS} =
S^i(A,x,Q^2,\vec{r},z) & = & 1 + N_{\rm WS}
[S^i(A,x,Q^2) - 1] \frac{\rho_A(\vec{r}, z)}{\rho_0} \label{wsparam} \, \, ,
\end{eqnarray}
where $N_{\rm WS}$ is chosen so that $(1/A) \int d^2r dz \rho_A(\vec{r}, z)
S^i_{\rm WS} = S^i$. If instead, shadowing arises due to multiple interactions
of the incident parton, shadowing should be proportional to the path length of
the parton through the nucleus,
\be
S^i_{\rm R}(A,x,Q^2,\vec{r},z) = \left\{ \begin{array}{ll}
 1 + N_{\rm R} (S^i(A,x,Q^2) - 1) \sqrt{1 - (r/R_A)^2} & 
\mbox{$r \leq R_A$} \label{rparam} \\
 1                                                        &
\mbox{$r > R_A$} \end{array} \right.
 \, \, .
\ee 
The normalization, $N_{\rm R}$, obtained after averaging over
$\rho_A(\vec{r},z)$, is slightly larger than $N_{\rm WS}$.  In both cases,  
at distances much greater than $R_A$, the nucleons
behave as free particles while in the center of the nucleus, the
modifications are larger than the average value $S^i$.  We have also studied
inhomogeneous shadowing assuming that shadowing is proportional to the path
length through the nucleus and find that
the central shadowing is somewhat
stronger than with $S_{\rm WS}^i$ in Eq.~(\ref{wsparam}).

\begin{figure}[htb]
\setlength{\epsfxsize=0.75\textwidth}
\setlength{\epsfysize=0.35\textheight}
\centerline{\epsffile{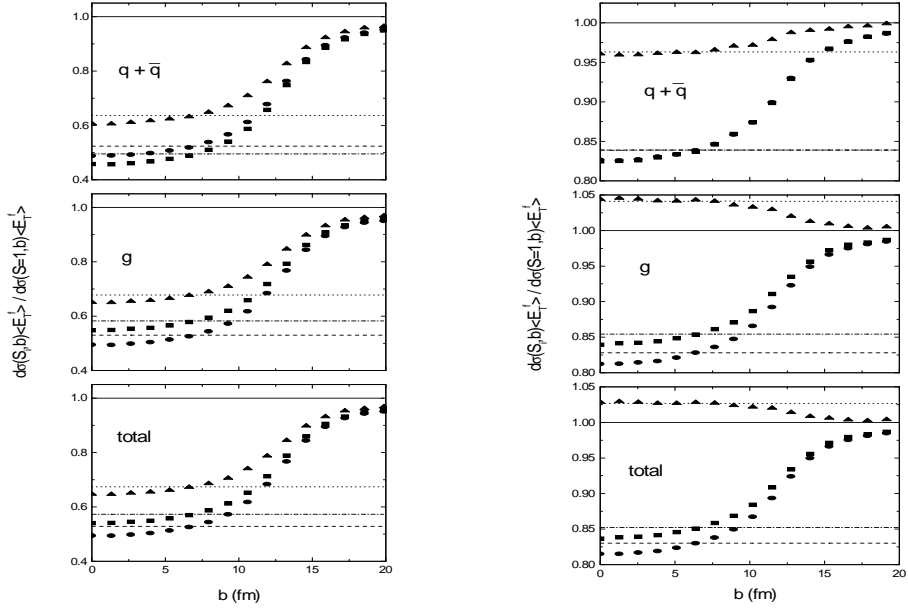}}
\caption[]{The impact parameter dependence of the first $E_T$ moment,
compared to that with
$S=1$ in CMS, $|y_{\rm max}| \leq 2.4$, at the LHC (left) and in STAR, $|y_{\rm
max}| \leq 0.9$, at RHIC (right) calculated with the GRV 94 LO
distributions and $p_0 =2$ GeV.  The upper plot shows the ratio for
quarks and antiquarks, the middle plot is the gluon ratio and the
lower plot is for the total.  The horizontal lines show the homogeneous
shadowing results: dashed for $S_1$, dot-dashed for $S_2$, and dotted
for $S_3$.  The inhomogeneous shadowing results using $S_{\rm WS}^i$
for $S_1$, circles,
$S_2$, squares, and $S_3$, triangles are also shown \cite{ekkv}.  }
\label{rat}
\end{figure}

The effect of the inhomogeneous shadowing, $S_{\rm WS}^i$ in
Eq.~(\ref{wsparam}), on $\overline E_T^f(b)$, 
calculated with the GRV 94 LO parton densities, is shown in
Fig.~\ref{rat} for CMS and STAR.  The CMS and ALICE ratios are similar as are
the STAR and PHENIX results. 
When $x$ lies in the shadowing region, central
collisions are more shadowed than the average.  In the antishadowing
region, central collisions are more antishadowed than the average.
When $b \sim R_A$, the homogeneous and inhomogeneous shadowing are
approximately equal.  When $b \sim 2R_A$, 
shadowing/antishadowing is significantly reduced.  As $b$
increases further, $S=1$ is approached asymptotically.  When inhomogeneous
shadowing is assumed to be proportional to $S_{\rm R}^i$, Eq.~(\ref{rparam}), 
we find
stronger shadowing at low impact parameter and a somewhat faster approach to
$S=1$ at large $b$.

\begin{figure}[htb]
\setlength{\epsfxsize=0.95\textwidth}
\setlength{\epsfysize=0.25\textheight}
\centerline{\epsffile{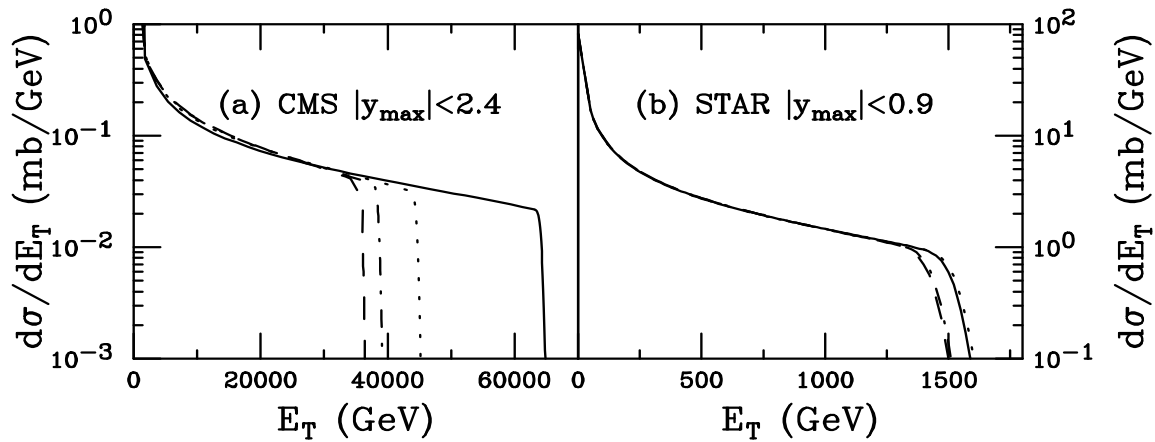}}
\caption[]{The $E_T$ distribution predicted for CMS, 
$|y_{\rm max}| \leq 2.4$, at the LHC and STAR, $|y_{\rm max}| \leq 0.9$,
at RHIC calculated with the GRV 94 LO distributions and $p_0=2$ GeV.
The homogeneous shadowing results for all $b$ are: no
shadowing (solid), $S_1$ (dashed), $S_2$ (dot-dashed), and
$S_3$ (dotted).  Modified from Ref.~\cite{ekkv}.
}
\label{etdist}
\end{figure}

We assume that the transverse energy distribution,
$d\sigma/dE_T$, can be
approximated by the Gaussian \cite{EKinit}
\begin{equation}
{d\sigma \over dE_T} = \int  {d^2b \over \sqrt{2\pi\sigma^{2}_E(b)}}
\exp{\bigg( \,- \, { [E_T-\overline E_T(b)]^2
\over 2\sigma^{2}_E(b)} \bigg)}, \, \, . \label{etgauss} \end{equation}
The total $E_T$ distribution is a convolution of the hard and soft
components with mean and standard deviation
\begin{eqnarray}
\overline E_T(b) = \sum_f \overline E_T^f (b)
+  T_{AB}(b) \epsilon_0 \,\,\,\,\,\,\,\,\,\,\,\,\,\,
\sigma^2_E(b) = \sum_f \overline{E_T^{2\ f}} -
\frac{\overline E_T^{H \, 2}}{\sigma^H(p_0)}
+ T_{AB}(b) \left( \epsilon_1-{\epsilon_0^2 \over \sigma^S_{pp}}\right) 
\label{stdave} \, \, .
\end{eqnarray}
The $E_T$ distributions
are shown for CMS and STAR in Fig.~\ref{etdist}.  At the LHC, 90\% of
the average $E_T$ comes from the hard component, causing the
maximum $E_T$ to be halved due to shadowing. In fact, the low $x$ gluon density
is high enough for the incoming gluon to have multiple hard collisions, up to
5 in central collisions with no shadowing, as it
traverses the nucleus when the GRV 94 LO distributions are used.  In contrast,
an incoming quark has at most 1 collision.  Similar
results with MRST LO distributions typically halve the hard $E_T$ 
and significantly reduce gluon multiple scattering.  Gluons
suffer 1.4-2 hard collisions while quarks scatter less than once \cite{ekkv}. 
At RHIC, since the
hard and soft components are comparable, the maximum $E_T$ is shifted
by only $\sim 7$\% when shadowing is included.  Indeed, for $S_3$, the
maximum $E_T$ is slightly increased. Thus shadowing significantly affects 
the initial conditions at LHC but
not at RHIC.  Ref.~\cite{ekkv} gives more details and discusses
inhomogeneous shadowing effects on $J/\psi$ and Drell-Yan production.

\end{document}